\documentclass[aoas,nameyear,dvips]{arximspdf}
\usepackage{shere}
\usepackage{graphicx}
\usepackage{breakurl}
\doi{10.1214/10-AOAS337}
\volume{4}
\issue{3}
\pubyear{2010}
\firstpage{1272}
\lastpage{1290}

\begin{document}
\begin{frontmatter}

\title{Topological inference for EEG and MEG\thanksref{M1}}
\runtitle{Topological inference}

\begin{aug}
\author{\fnms{James M.} \snm{Kilner}} \and
\author{\fnms{Karl J.} \snm{Friston}\corref{}\ead[label=e1]{k.friston@fil.ion.ucl.ac.uk}}
\runauthor{J. M. Kilner and K. J. Friston}
\affiliation{University College London}
\thankstext{M1}{Supported by The Wellcome Trust.}
\address{Institute of Neurology\\
University College London\\
The Wellcome Trust Centre for Neuroimaging\\
12 Queen Square\\
London WC1N 3BG\\
United Kingdom\\
\printead{e1}} 
\end{aug}

\received{\smonth{8} \syear{2009}}
\revised{\smonth{1} \syear{2010}}

\begin{abstract}
Neuroimaging produces data that are continuous in one or more
dimensions. This calls for an inference framework that can handle data
that approximate functions of space, for example, anatomical images,
time--frequency maps and distributed source reconstructions of
electromagnetic recordings over time. Statistical parametric mapping
(SPM) is the standard framework for whole-brain inference in
neuroimaging: SPM uses random field theory to furnish $p$-values that
are adjusted to control family-wise error or false discovery rates, when
making topological inferences over large volumes of space. Random field
theory regards data as realizations of a continuous process in one or
more dimensions. This contrasts with classical approaches like the
Bonferroni correction, which consider images as collections of discrete
samples with no continuity properties (i.e., the probabilistic behavior
at one point in the image does not depend on other points). Here, we
illustrate how random field theory can be applied to data that vary as a
function of time, space or frequency. We emphasize how topological
inference of this sort is invariant to the geometry of the manifolds on
which data are sampled. This is particularly useful in electromagnetic
studies that often deal with very smooth data on scalp or cortical
meshes. This application illustrates the versatility and simplicity of
random field theory and the seminal contributions of Keith Worsley
(1951--2009), a key architect of topological inference.
\end{abstract}


\end{frontmatter}

\section{Introduction} \label{sec:1}

This paper is about inferring treatment effects or responses that are
expressed in image data. The~problem we consider is how to accommodate
the multiplicity of data and correlations due to smoothness, when
adjusting for the implicit multiple comparison problem. The~data we have
in mind here are images that can be treated as discrete samples from a
function of some underlying support: for example, two-dimensional images
of the brain sampled from evenly spaced points (i.e., a grid) in
anatomical space. In brief, the multiple comparison problem can be
dissolved by modeling the data as samples from random fields with known
(or estimable) covariance functions over their support. This allows one
to use results from random field theory to determine the topological
behavior (e.g., the number of peaks above some threshold) of summary
statistic images, under the null hypothesis. Because we treat the data
and derived statistical processes as implicit functions of some metric
space, this approach is closely related to functional data analysis
[Ramsay and Silverman (\citeyear{RaSi1997})]. When this functional perspective is
combined with random field theory (as a probabilistic model of data), we
get a generic inference framework (topological inference) that is used
widely in brain mapping and other imaging fields.

We will review topological inference in neuroimaging with a special
focus on electromagnetic (EEG and MEG) data. In particular, we stress
the generality of this approach and show that random theory can be
applied to data-features commonly used in EEG and MEG. These data
include interpolated scalp-maps, time--frequency maps of single-channel
data, cortically constrained maps of current density, source
reconstruction on the cortex or brain volume, etc. Irrespective
of the underlying geometry or support of these data, the topological
behavior of their associated statistical parametric maps is invariant.
This means one can apply established procedures directly to make
inferences about evoked and induced responses in sensor or source-space.
This reflects the simplicity and generality of topological inference and
provides a nice vehicle to illustrate the seminal work of Keith Worsley,
who sadly passed away shortly before this article was written.

Conventional whole-brain neuroimaging data analysis uses some form of
statistical parametric mapping. This entails a parametric model (usually
a general linear model) of data at each point in image space to produce
a statistical parametric map (usually of a Student's $t$-statistic).
Topological inference about regional effects then uses random field
theory to control for the implicit multiple comparisons problem. This is
standard practice in imaging modalities like functional magnetic
resonance imaging (fMRI) and positron emission tomography. However, the
application to electroencephalography (EEG) and magnetoencephalography
(MEG) data is relatively new. An interesting feature of electromagnetic
data is that they are often formulated on meshes or manifolds, which may
seem to complicate the application of random field theory. This paper
shows that random field theory can be applied directly to
electromagnetic data on manifolds and that it accommodates the
anisotropic and complicated spatial dependencies associated with smooth
electromagnetic data-features. We have chosen to illustrate topological
inference on electromagnetic data-features because they are inherently
smooth and show profound spatial dependencies, which preclude classical
procedures. These dependencies arise from preprocessing steps (e.g.,
interpolation to produce scalp maps or source reconstruction, under
regularizing smoothness constraints) or from the nature of the
data-features per se (e.g., physiological smoothness in
time-series or time--frequency smoothness induced by wavelet
decomposition).

EEG and MEG are related noninvasive neuroimaging techniques that provide
measures of human cortical activity. EEG and MEG typically produce a
time-varying modulation of signal amplitude or frequency-specific power
in some peristimulus period, at each electrode or sensor. The~majority
of researchers are interested in whether condition-specific effects
(observed at particular sensors and peristimulus times) are
statistically significant. However, this inference must correct for the
number of statistical tests performed. In other words, the family-wise
error (FWE) rate should be controlled. For independent observations, the
FWE rate scales with number of observations. A simple but inexact method
for controlling FWE is a Bonferroni correction. However, this procedure
is rarely adopted in neuroimaging because it assumes that neighboring
observations are independent: when there is a high degree of correlation
among neighboring samples (e.g., when data-features are smooth), the
correction is far too conservative.

Although the multiple comparisons problem has always existed for\break EEG/MEG
analyses (due to the number of time bins in the peristimulus time
window), the problem has become more acute with the advent of
high-density EEG-caps and MEG sensor arrays that increase the number of
observations across the scalp. In many analyses, the multiple
comparisons problem is circumvented by restricting the search-space
prior to inference, so that there is only one test per repeated measure.
This is usually accomplished by averaging the data over pre-specified
sensors and time-bins of interest. This produces one summary statistic
per subject per condition. In many instances, this is a powerful and
valid way to side-step the multiple comparisons problem; however, it
requires the space-of-interest be specified a priori. A
principled specification of this space could use orthogonal or
independent data-features. For example, if one was interested in the
attentional modulation of the N170 (a typical event-related wave
recorded 170 ms after face presentation), one could first define the
electrodes and time-bins that expressed a N170 (compared to baseline)
and then test for the effects of attention on their average. Note that
this approach assumes that condition-specific effects occur at the same
sensors and time and is only valid when selection is not biased [see
Howell (\citeyear{Ho1997}); Kriegeskorte et al. (\citeyear{KrSiBeBa2009})]. In situations where the
location of evoked or induced responses is not known a priori,
or cannot be localized independently, one can use topological inference
to search over some space for significant responses; this is the
approach we consider.

This paper comprises two sections. The~first reviews the application of
random-field theory [RFT; Worsley et al. (\citeyear{WoEvMaNe1992}, \citeyear{WoMaNeNaFrEv1996})] to statistical
parametric maps [SPMs; Friston et al. (\citeyear{FrFrLiFr1991}, \citeyear{FrWoFrMaEv1994})] of MEG/EEG data
over space, time and frequency. In the second section we illustrate the
basic procedures by applying RFT to SPMs of MEG/EEG data and try to
highlight the generality of the approach. In this article we focus on
FWE but note the same topological thinking and random field theory
results can also be used to control false discovery rate [FDR; Benjamini
and Hochberg (\citeyear{BeHo1995})]. We provide a brief review of Topological FDR for
image analysis in the discussion.

\section{Random fields and topological inference} \label{sec:2}

In this section we review RFT and its central role in statistical
parametric mapping. We first provide a heuristic overview and then give
details, with a special focus on issues that relate to EEG/MEG analyses.
RFT provides an established method for assigning $p$-values to
topological features of SPMs in the analysis of functional magnetic
resonance (fMRI) and other anatomical, metabolic or hemodynamic images.
More recently, it has been applied to hierarchical models of EEG/MEG
data [Park et al. (\citeyear{PaKwYoPaKiKiHa2002}); Barnes and Hillebrand (\citeyear{BaHi2003}); Kiebel and
Friston (\citeyear{KiFr2004}); Henson et al. (\citeyear{HeMaSiBaHiFr2007}); Garrido et al. (\citeyear{GaFrKiStBaKi2008})], global
field power statistics [Carbonell et al. (\citeyear{CaGaVaWoBiDaBoPa2004})], time--frequency data
[Kilner,  Kiebel and Friston (\citeyear{KiKiFr2005})], current source density maps [Pantazis et al.
(\citeyear{PaNiBaLe2005})] and even frequency by frequency coupling maps from dynamic
causal modeling [Chen, Kiebel and Friston (\citeyear{ChKiFr2008})].

\subsection{Statistical parametric mapping} \label{sec:2.1}

Statistical parametric maps (e.g.,\break  \mbox{$t$-maps}) are fields with values that
are, under the null hypothesis, distributed according to a known
probability distribution. This is usually the Student's $t$- or
$F$-distributions. SPMs are interpreted as continuous statistical
processes by referring to the probabilistic behavior of random fields
[Worsley et al. (\citeyear{WoEvMaNe1992}, \citeyear{WoMaNeNaFrEv1996}); Friston et al. (\citeyear{FrFrLiFr1991}, \citeyear{FrHoWoPoFrFr1995})]. Usually, a
general linear model is used to estimate the parameters that best
explain some data-features. One fits a general linear model at each
point (vertex or voxel) of the search-space and computes the usual
statistics [see Friston et al. (\citeyear{FrHoWoPoFrFr1995}) for details]; these constitute the
SPM. The~search-space can, in principle, be of any dimensionality and
could be embedded in a higher dimensional space. RFT is then used to
resolve the multiple comparisons problem that occurs when making
inferences over the search-space: Adjusted $p$-values are obtained by
using results for the expected Euler characteristic of the excursion set
of a smooth statistical process.

The~Euler characteristic (or Euler--Poincar\'{e} characteristic) is a
topological invariant that describes the shape or structure of a
manifold, regardless of the way it is stretched or distorted. It was
defined classically for the surfaces of polyhedra, where it is simply
the number of faces and corners, minus the number of edges. In our
context, it effectively counts the number of connected regions (minus
the number of holes) in the excursion set that remains after thresholding
an SPM. At very high thresholds the Euler characteristic (abbreviated
here ``EC'') basically reduces to the number of suprathreshold peaks and
the expected EC becomes the probability of getting a peak above
threshold by chance (under the Poisson clumping heuristic).

The~expected EC therefore approximates the probability that the SPM
exceeds some height by chance. This is the same as the $p$-value based
on the null distribution of the maximum statistic over search-space. The~ensuing $p$-values can be used to find a corrected height threshold or
assign a corrected $p$-value to any observed peak in the SPM [see  Worsley (\citeyear{Wo2007}) for an introduction to RFT]. The~fundamental
advantage of RFT is that it models continuous statistical processes and
not a collection of individual statistics. This means that RFT can be
used to characterize topological features of the SPM like peaks. The~key
intuition behind RFT procedures is that they control the false positive
rate of topological features, not the tests themselves. By way of
contrast, a Bonferroni correction controls the false positive rate of
tests (at vertices, time--frequency bins or voxels), which would be
unnecessarily conservative when the data are smooth. RFT has become a
cornerstone of inference in human brain mapping that enables researchers
to adjust their $p$-values to control false positive rates over many
different sorts of search-spaces with spatial dependencies.

\subsection{Random field theory} \label{sec:2.2}

The~assumptions under which the random field correction operates are
quite simple and are satisfied by high-density EEG/MEG data because of
their inherent smoothness in space and time. As noted above, the key
null distribution is that of the maximum statistic over the search
volume. By evaluating any observed statistic, in relation to the null
distribution of its maximum, one is implicitly implementing a multiple
comparisons procedure for continuous data. An analytic form of this
distribution is derived using results from RFT. These results use the
expected Euler characteristic of excursion sets above some specified
threshold. For high thresholds this expectation is the same as the
probability of getting a maximum statistic above threshold. By treating
the data, under the null hypothesis, as continuous random fields, the
distribution of the Euler characteristic of any statistical process
derived from these fields can be used as an approximation to the null
distribution required for inference. When using a general linear model
the random (component) fields correspond to error fields. RFT assumes
that these are a good lattice approximation of an underlying random
field. Furthermore, the expressions require that the error fields are
multivariate Gaussian with a differentiable autocorrelation function. It
is a common misconception that this correlation function has to be
Gaussian: it does not. Furthermore, the autocorrelation function does
not have to be stationary or isotropic. The~ensuing $p$-value is a
function of the search volume, over any arbitrary number of dimensions,
and the local smoothness of the underlying error fields, which can be
expressed in terms of full-width, half-maximum (FWHM). A useful concept
that combines these two aspects of the search-space is the number of
``resolution elements'' (\textit{resels}---see below). The~resel count
corresponds to the number of FWHM elements that comprise the search
volume. Heuristically, these encode the number of independent
observations. In other words, even a large search volume may contain a
relatively small number of resels, if it is smooth. This calls for a
much less severe adjustment to the $p$-value than would be obtained with
a Bonferroni correction based on the number of bins, voxels or vertices.

We now develop these intuitions more formally, with a didactic summary
of random field theory based on Taylor and Worsley (\citeyear{TaWo2007}) and implemented
in conventional software such as \textit{fMRIStat}, \textit{SurfStat}
and \textit{SPM8}. The~associated software is available from
\url{http://www.math.mcgill.ca/keith/fmristat/},
\url{http://www.math.mcgill.ca/keith/surfstat/} and
\url{http://www.fil.ion.ucl.ac.uk/spm/}.

\subsection{The~Euler characteristic} \label{sec:2.3}

Imagine that we have collected some\break  EEG/MEG data and have interpolated
them to produce a 2D scalp-map of responses at one point in peristimulus
time for two conditions and several subjects. We now compare the two
conditions with a statistical test; such that we now have $T$
(test)-values at every vertex in our scalp-map. We are interested in
the \textit{T}-value, above which we can declare that differences are
significant at some $p$-value that is adjusted for FWE.

RFT gives adjusted $p$-values by using results for the expected Euler
characteristic (EC) of the excursion set of a smooth statistical
process. The~expected EC of the excursion set is an accurate
approximation to the $p$-value of the maximum of a smooth, nonisotropic
random field or SPM of some statistic $T(s)$ at Euclidean coordinates $s
\in S$, above a high threshold $t$ and is given by
\begin{equation}\label{for1}
P\Bigl( \mathop{\max}  _{s \in S}T(s) \ge t \Bigr) = \sum_{d = 0}^{D}
\ell_{d}(S,\Lambda)\rho_{d}(t),
\end{equation}
where $\ell_{d}(S,\Lambda)$ are the \textit{Lipschitz--Killing
curvatures} (LKC), of the\break $D$-dimensional search-space $S
\subset\Re^{D}$, and $\rho_{d}(u)$ are the \textit{EC densities}. These
are two important quantities: put simply, the LKC measures the
topologically invariant ``volume'' of the search-space. In other words, it
is a measure of the manifold or support of the statistical process that
does not change if we stretch or distort it. The~EC density is the
corresponding ``concentration'' of events (excursions or peaks) we are
interested in. Effectively, the product of the two is the number of
events one would expect by chance (the expected Euler characteristic).
When this number is small, it serves as our nominal false positive rate
or $p$-value.

Equation (\ref{for1}) shows that the $p$-value receives contributions from all
dimensions of the search-space, where the largest contribution is
generally from the highest dimension ($D$). In the example above, we had
a two-dimensional $D = 2$ search-space. Each contribution comprises two
terms: (i) The~LKC, which measures the effective volume, after
accounting for nonisotropic smoothness in the component or error fields.
This term depends on the geometry of the search-space and the smoothness
of the errors but not the statistic or threshold used for inference.
(ii)~Conversely, the EC density, which is the expected number of
threshold excursions per LKC measure, depends on the statistic and
threshold but not the geometry or smoothness. Closed-form expressions
for the EC density are available for all statistics in common use [see
Worsley et al. (\citeyear{WoLiAsPeDuMoEv2002})].

\subsection{Smoothness and resels} \label{sec:2.4}

The~LKC encodes information about the support and local correlation
function of the underlying error fields $Z(s)$. The~correlation structure
is specified by their roughness or the variability of their gradients,
$\dot{Z}(s)$, at each coordinate
\begin{equation}\label{for2}
\Lambda(s) = \operatorname{Var}(\dot{Z}(s)).
\end{equation}
In the isotropic case, when the correlations are uniform, $\Lambda(s) =
I_{D \times D}$, the LKC reduces to \textit{intrinsic volume}
\begin{equation}\label{for3}
\ell_{d}(S,I_{D \times D}) = \mu_{d}(S).
\end{equation}
The~intrinsic volume is closely related to the intuitive notion of a
volume and can be evaluated for any regular manifold (or computed
numerically given a set of vertices and edges defining the
search-space). Note that when $S$ is a $D$-dimensional manifold embedded
in a higher dimensional space, the higher dimensional volumes are all
zero, so that the sum in equation (\ref{for1}) need only go to the dimensionality
of the manifold, rather than the dimensionality of the embedding space.
In the example above, we can think of our 2D scalp map embedded in a 3D
head-space; however, we only need consider the two dimensions of the
scalp-map or manifold. The~LKC term that makes the largest contribution
to the $p$-value is the final volume term
\begin{equation}\label{for4}
\ell_{D}(S,\Lambda) = \int_{S} |\Lambda(s)|^{1/2} \,ds = (4\ln
2)^{D/2}\operatorname{resels}_{D}(S).
\end{equation}
This generalization of the LKC is the resel count [Worsley et al.
(\citeyear{WoMaNeNaFrEv1996})], which reflects the number of effectively independent
observations. It can be seen from equation (\ref{for4}) that the resels (or LKC)
increase with both volume and roughness.

\subsection{Estimating the resel count} \label{sec:2.5}

The~resel count can be estimated by replacing the coordinates $s \in S$
by normalized error fields $u(s) \in\Re^{n}$, to create a new space
\begin{equation}\label{for5}
u(s) = \frac{Z(s)}{\sqrt{n}} \approx\frac{r(s)}{\Vert r(s) \Vert}.
\end{equation}
Here, $r(s)$ are $n$ normalized residual fields from our general linear
model. Crucially, the intrinsic volume at any point in this new space is
the LKC
\begin{equation}\label{for6}
\ell_{d}(S,\Lambda) = \mu_{d}(u(s)).
\end{equation}
This elegant device was proposed by Worsley et al. (\citeyear{WoAnKoMaEv1999}). It says that
to estimate the LKC, one simply replaces the Euclidean coordinates by
the normalized residuals, and proceeds as if $u(s)$ were isotropic. The~basic idea is that $u(s)$ can be thought of as an estimator of $S$ in
isotropic space, in the sense that the local geometry of $u(s)$ is the
same as the local geometry of $S$, relative to $Z(s)$ [Taylor and
Worsley (\citeyear{TaWo2007})]. This equivalence leads to the following estimator:
\begin{equation}\label{for7}
\ell_{D}(S,\Lambda) = \frac{1}{D!}\sum_{i = 1}^{N} |\Delta
u_{i}^{T}\Delta u_{i}|^{1/2} ,
\end{equation}
where $\Delta u = [\Delta u_{1}, \ldots,\Delta u_{D}]$ are the finite
differences between neighboring vertices of the $N$ components that tile
the search manifold (e.g., edges, triangles, tetrahedra, etc.),
note that this approximation does not depend on the Euclidean
coordinates of the vertices, only how they are connected to form
components [Worsley et al. (\citeyear{WoAnKoMaEv1999})]. At present, the \textit{SPM8}
software (but not \textit{SurfStat}) uses the following LKC estimator,
\begin{equation}\label{for8}
\ell_{d}(S,\Lambda) = \mu_{d}(S)\biggl(\frac{\ell _{D}(S,\Lambda )}{\mu
_{D}(S)}\biggr)^{d/D},
\end{equation}
and the approximation, $\Delta u_{i}^{T}\Delta u_{i} \approx\Delta
r_{i}^{T}\Delta r_{i} / r_{i}^{T}r_{i}$, which assumes $\Vert  r_{i} \Vert
\approx\Vert  r_{j} \Vert $ for connected vertices; this is generally true,
provided the error variance changes sufficiently smoothly.

The~important result above is equation (\ref{for6}), which allows one to estimate
the intrinsic volume of $u(s)$, which is the LKC of $S$. However, this is
another perspective on equation (\ref{for7}) that comes from an estimator based
on equation (\ref{for4}) [see Kiebel et al. (\citeyear{KiPoFrHoWo1999})],
\begin{equation}\label{for9}
\ell_{D}(S,\Lambda) = \sum_{i = 1}^{N} |\Lambda(s_{i})|^{1/2} \Delta
S_{i}.
\end{equation}
Comparison with equation (\ref{for7}) suggests that the determinant of finite
differences can also be regarded as an estimate of the local roughness
times the volume $\Delta S_{i}$ of the $i$th component of search-space,
\begin{equation}\label{for10}
|\Lambda(s_{i})|^{1/2}\Delta S_{i} = \frac{1}{D!}|\Delta
u_{i}^{T}\Delta u_{i}|^{1/2}.
\end{equation}
Effectively, the dependency of the local LKC on volume and the distances
between vertices (implicit in evaluating the gradients) cancel, so that
we need only consider the finite differences. \textit{In short, the
geometry encoded in the geodesic distances among vertices (or voxels)
has no effect on the LKC or ensuing $p$-values.} This means we can take
any nonisotropic statistical field defined on any $D$-dimensional
manifold embedded in a high-dimensional space (e.g., a cortical mesh in
anatomical space) and treat is as an isotropic $D$-dimensional SPM,
provided we replace the gradients of the normalized residuals (which
depend on the geometry) with finite differences among connected vertices
(which do not). This invariant aspect of the resel count (or the LKC)
estimator speaks to the topological nature of inference under random
field theory. This is summarized nicely in Taylor and Worsley (\citeyear{TaWo2007}):

\begin{quote}
``Note first that the domain of the random fields could be
warped or deformed by a one-to-one smooth transform without
fundamentally changing the nature of the problem. For example, we could
`inflate' the average cortical surface to a sphere and carry out all our
analysis on the sphere. Or we could use any convenient shape: the
maximum of the Student's $t$-statistic would be unchanged, and so would
the Euler characteristic of the excursion set. Of course the correlation
structure would change, but then so would the search region, in such a
way that the effects of these on the LKC, and hence the expected Euler
characteristic, cancel.''
\end{quote}

\noindent The~fact that the resels are themselves a topological measure is
particularly important for EEG and MEG data. This means that the resel
count does not depend on the Euclidean coordinates or geometry of the
data support. In other words, one can take data from the vertices of a
cortical mesh embedded in a 3D space and treat it as though it came from
a flat surface. This ability to handle nonisotropic correlations on
manifolds with an arbitrary geometry reflects the topographic nature of
RFT and may find a particularly powerful application in EEG and MEG
research. In the next section we demonstrate the nature of this
application.

\section{Illustrative applications} \label{sec:3}

In this section we illustrate RFT, as implemented in \textit{SPM8}, to
adjust $p$-values from SPMs of space-time MEG data. Data were recorded
from 14 subjects (9 males, age range 25--45 yrs). All subjects gave
informed written consent prior to testing under local ethical committee
approval. MEG was recorded using 275 third-order axial gradiometers with
the Omega275 CTF MEG system (VSMmedtech, Vancouver, Canada) at a
sampling rate of 480~Hz. Details of the experimental design will be
found in Kilner, Marchant and Frith (\citeyear{KlMaFr2006}). Here, we describe the features of the
task that are relevant for the analyses used in this paper; namely,
event-related analyses of right-handed button presses in the time and
frequency domains.

Subjects performed four sessions of a task consecutively. In each
session, subjects performed forty button presses with their right index
finger, giving a total of 160 trials. All MEG analyses were performed in
\textit{SPM8}: First, the data were epoched relative to the button
press. The~data were band-pass filtered between 0.1 and 45 Hz using a
time window of $-$500 to 1000 ms and down-sampled to 100 Hz. For
event-related field (ERF) analyses, the data were averaged across trials
for each sensor. For time--frequency (TF) analyses, induced oscillations
were quantified using a (complex Morlet) wavelet decomposition of the
MEG signal, over a 1--45 Hz frequency range. The~wavelet decomposition
was performed for each trial, sensor and subject. The~ensuing
time--frequency maps were averaged across trials. For the purposes of
this paper, we were interested in demonstrating significant ``rebound''
effects in the 15--30 Hz range [Salmelin and Hari (\citeyear{SaHa1994})]. Therefore, the
time--frequency maps were averaged across the 15--30 Hz frequency band to
produce a time-varying modulation of the so-called \textit{Beta} power
at each sensor.

\begin{figure}

\includegraphics{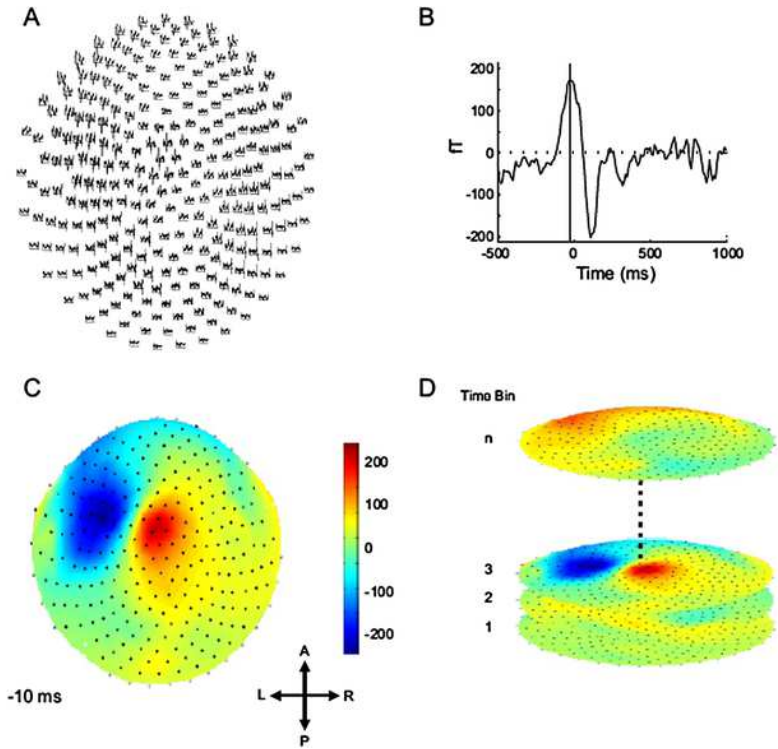}

\caption{Single-subject ERF data. (\textup{A}) shows the average ERF for a single
subject. The~data are plotted for 275 sensors across peristimulus time.
(\textup{B}) shows the average ERF for the same subject from one sensor.
The~vertical line indicates the maximum positive value of this ERF. (\textup{C}) shows
the sensor-space interpolated map across all sensors at $-10$~ms,
indicated by the line in (\textup{B}). (\textup{D}) shows how the 3D sensor-space-time data
volume is formed.}\label{fig1}
\end{figure}

For both the ERF and TF analyses, the sensor-data at each time bin were
interpolated to produce a 2D sensor-space map on a $64\times 64$ mesh aligned to
the left--right and anterior--posterior axes [e.g., Figure \ref{fig1}(C)]. A 3D
data-array was generated for each subject by stacking these scalp-maps
over peristimulus time [Figure~\ref{fig1}(D)]. This produces a 3D image, where the
dimensions are space (left--right and anterior--posterior) and time. For
each subject, a second reference 3D image was generated that was the
mean amplitude of the signal at each sensor, replicated at each time
point. These space-time maps were smoothed using a Gaussian kernel (FWHM
$6 \times  6$ spatial bins and 60 ms) prior to analysis. This smoothing step is
essential. First, it assures the assumptions of RFT are not violated.
These assumptions are that the error fields conform to a good lattice
approximation of a random field with a multivariate Gaussian
distribution. Second, it blurs effects that are focal in space or time,
ensuring overlap among subjects. It should be noted that although
smoothing is an important pre-processing procedure, it is not an
inherent part of topological inference: RFT estimates the smoothness
directly from the (normalized residual) data, during the estimation of
the resel count. This means one has the latitude to smooth in a way that
emphasizes the data-features of interest. For example, with cortical or
scalp manifolds one might use weighted [e.g., Pantazis et al. (\citeyear{PaNiBaLe2005})] or
un-weighted graph-Laplacian operators to smooth the data on their meshes
[see Harrison et al. (\citeyear{HaPeDaFr2008}) for a fuller discussion]. An un-weighted
graph-Laplacian produces the same smoothing as convolution with a
Gaussian kernel on a regular grid: this is the approach used here.

We then generated SPMs on a regular 3D grid by performing a series of
$t$-tests comparing the response to the mean image at every bin in
scalp-space and time. This is called a mass-univariate approach and is
identical to that adopted in the analysis of fMRI data.

\subsection{Event-related field analysis} \label{sec:3.1}

The~analysis of the ERF is typical of high-density EEG/MEG studies.
Figure~\ref{fig1} illustrates the problem of making a statistical inference on
such multidimensional data. The~data for each subject consists of 151
observations across time at 275 observations in sensor-space [Figure~\ref{fig1}(A)]. Figure~\ref{fig1}(B) shows the time-course of the ERF at a sensor that
evidences a movement-evoked field [e.g., Hari and Imada (\citeyear{HaIm1999})]. Note
that the early onset is due to alignment to the button press and not the
onset of movement. If one looks at the modulation of this signal across
space in the sensor-space map (at the maximum of the effect), a clear
dipole field pattern can be seen [Figure~\ref{fig1}(C)].

\begin{figure}

\includegraphics{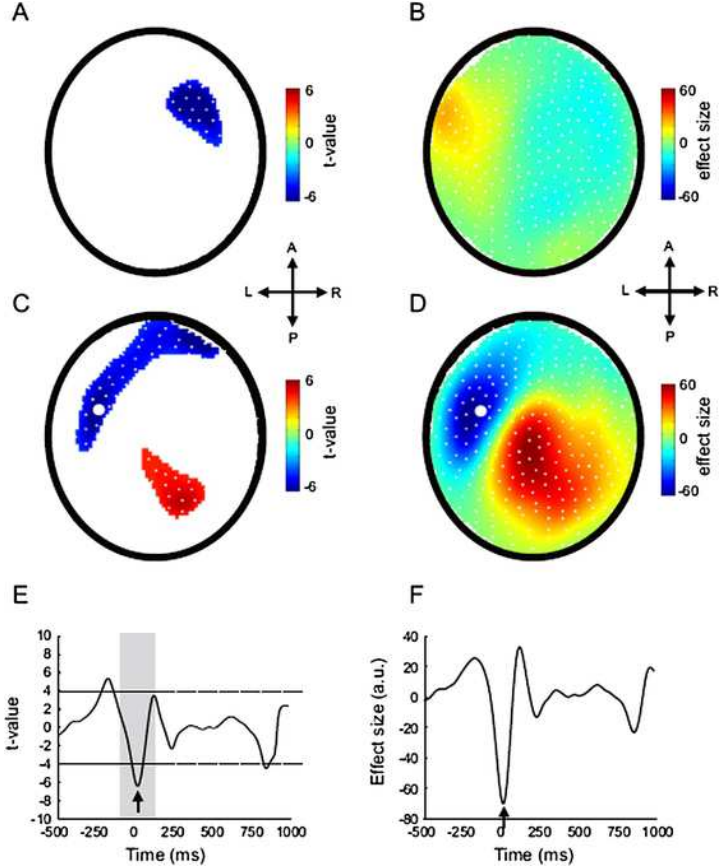}

\caption{SPM analysis of movement ERF. (\textup{A}) shows the
$\operatorname{SPM}(t)$, thresholded
at $p<0.001$ (uncorrected), showing where the effects were less than the
mean. The~peak value within this cluster is significant at $p<0.05$ (FWE
corrected). (\textup{B}) shows the sensor-space map of (nonstandardized) effect
size across all sensors at the time where the SPM was maximal. The~effect
size is proportional to the grand mean across subjects. (\textup{C}) shows
the SPM, thresholded at $p<0.001$ (uncorrected), showing where the effects
were greater and less than the mean. The~sensor within these clusters
that was significant at $p<0.05$ (corrected for a small search volume) is
shown by the white circle. (\textup{D}) shows the sensor-space map of effect size
across all sensors at the time point where the SPM in (\textup{C})
was maximal. (\textup{E})
shows the time course of the SPM from the sensor shown in white in (\textup{C}).
The~dashed lines show the uncorrected threshold for the Student's
$t$-statistic at $p<0.001$. (\textup{F}) shows the corresponding plot for the effect
size. In both (\textup{E}) and (\textup{F}) the arrow shows the time where the SPM was
maximal.}\label{fig2}
\end{figure}

We now want to test for responses over space and time. An SPM for
effects greater or less than the mean was calculated using a paired
$t$-test over subjects. In this example, only one peak was greater than
a threshold adjusted for the entire search volume [$p<0.05$ corrected;
Figure~\ref{fig2}(A)]. The~peak value occurred 120 ms prior to the button press and
was within a cluster of right frontal sensors [Figure~\ref{fig2}(A) and Table \ref{tab1}].
Figure~\ref{fig2}(B) shows the average (nonstandardized) effect size across
sensor-space at $-$120~ms: when comparing the thresholded SPM
[Figure~\ref{fig2}(A)]
and the effect-size map [Figure~\ref{fig2}(B)], it can be seen that the peaks of
the SPM and effect-size are in different places. There is no reason why
they should be in the same location, because the Student's $t$-statistic
reflects the effect-size and standard error. This example highlights the
benefit of inference that is controlled for FWE across space and time,
namely, that one can discover effects that were not predicted a priori. However, it also suggests significant effects should be
interpreted in conjunction with the effect-size. In other words,
although the peak in the SPM tells us where differences are significant,
it does not necessarily identify the maximum response in a quantitative
sense.

In most instances, searches over SPMs are constrained or directed. This
is common in fMRI when we know a priori where in the brain to
look. The~same is true for EEG/MEG data. For the example considered
here, we may want to constrain the search-space to some peristimulus
time window. However, in contradistinction to conventional approaches,
we do not average over the volume of interest space but use it to
constrain the search and increase its sensitivity. In this instance, the
RFT adjusts $p$-values over a smaller volume and implements a less
severe adjustment. For the ERF shown here, given the previous
literature, we defined a time-window of interest of 200 ms, starting at
$-$100 ms before the button press. Within this time-window, the peak of
the SPM occurred at 10 ms at central sensors overlying the left
hemisphere and was significant at $p<0.05$ [corrected: Figure~\ref{fig2}(C) and
Table~\ref{tab2}]. Figure~\ref{fig2}(C) also shows the thresholded SPM at $p<0.001$
(uncorrected) for the opposite contrast, where responses were greater
than the mean. When comparing this thresholded SPM image to the
corresponding effect-size image [Figure~\ref{fig2}(D)], one can see that the
sensors that survive the threshold in Figure~\ref{fig2}(C) display a classical
single-dipole field pattern [Figure~\ref{fig2}(D)].

\begin{table}
\caption{Statistical results of a full volume analysis in \textup{SPM8}.
The~table entries (from left to right) represent the following: the
adjusted or corrected $p$-value based on random field theory that
controls false positive rate; the equivalent $p$-value ($q$-value)
controlling false discovery rate; the maximum Student's $t$-statistic;
its $Z$-score equivalent; its uncorrected $p$-value; the time at which
this peak occurred. The~footnotes provide details of the search volume
and topological features expected under the null hypothesis (see
\protect\url{http://www.fil.ion.ucl.ac.uk/spm/} for details)}
\label{tab1}
\begin{tabular*}{\textwidth}{@{\extracolsep{\fill}}lccccc@{}}
\hline
\multicolumn{6}{@{}c@{}}{\textbf{Peak level}} \\
\hline
$\bolds{p}_\mathbf{FWE\mbox{\textbf{-}}corr}$ & $\bolds{p}_\mathbf{FDR\mbox{\textbf{-}}corr}$ & $\bolds{t}$ & $\bolds{Z}$ & $\bolds{p}_\mathbf{uncorrected}$ & \textbf{Time (ms)}\\
\hline
0.036 & 0.017 & 8.71 & 4.80 & 0.000 & $-$120\\
\hline
\end{tabular*}
\tabnotetext[]{tm}{Statistics: $p$-values restricted to the entire search volume.
Height threshold: $T = 3.93$, $p = 0.001$ (0.993); Degrees of freedom $= [1.0,12.0]$;
Extent threshold: $k = 0$ bins, $p = 1.000$ (0.993); Smoothness $\mathrm{FWHM} = 13.1$ 17.5 8.6 \{bins\};
Expected bins per cluster, $\langle k\rangle  = 121.669$; Search vol.: 1,808,083 bins; 230.3 resels;
Expected number of clusters, $\langle c\rangle = 4.96$;
Expected false discovery rate, ${\leq}0.03$.}
\end{table}

\begin{table}[b]
\caption{Statistical results of a small volume analysis in
\textup{SPM8}. This table uses the same format as Table \protect\ref{tab1}}
\label{tab2}
\begin{tabular*}{\textwidth}{@{\extracolsep{\fill}}lccccc@{}}
\hline
\multicolumn{6}{@{}c@{}}{\textbf{Peak level}} \\
\hline
$\bolds{p}_\mathbf{FWE\mbox{\textbf{-}}corr}$ & $\bolds{p}_\mathbf{FDR\mbox{\textbf{-}}corr}$ & $\bolds{t}$ & $\bolds{Z}$ & $\bolds{p}_\mathbf{uncorrected}$ & \textbf{Time (ms)}\\
\hline
0.013 & 0.007 & 6.86 & 4.29 & 0.000 & 10\\
\hline
\end{tabular*}
\tabnotetext[]{tm}{Statistics: $p$-values restricted to $-$100--100 ms.
Height threshold: $T = 3.93$, $p = 0.001$ (0.289); Degrees of freedom $= [1.0, 12.0]$;
Extent threshold: $k = 0$ bins, $p = 1.000$ (0.289); Smoothness $\mathrm{FWHM} = 13.1$ 17.5 8.6 \{bins\};
Expected bins per cluster, $\langle k\rangle = 121.669$; Search vol.: 82,340 bins; 11.5 resels;
Expected number of clusters, $\langle c\rangle  = 0.34$;
Expected false discovery rate, ${\leq}0.01$.}
\end{table}

\subsection{Time--frequency analysis} \label{sec:3.2}

\begin{figure}

\includegraphics{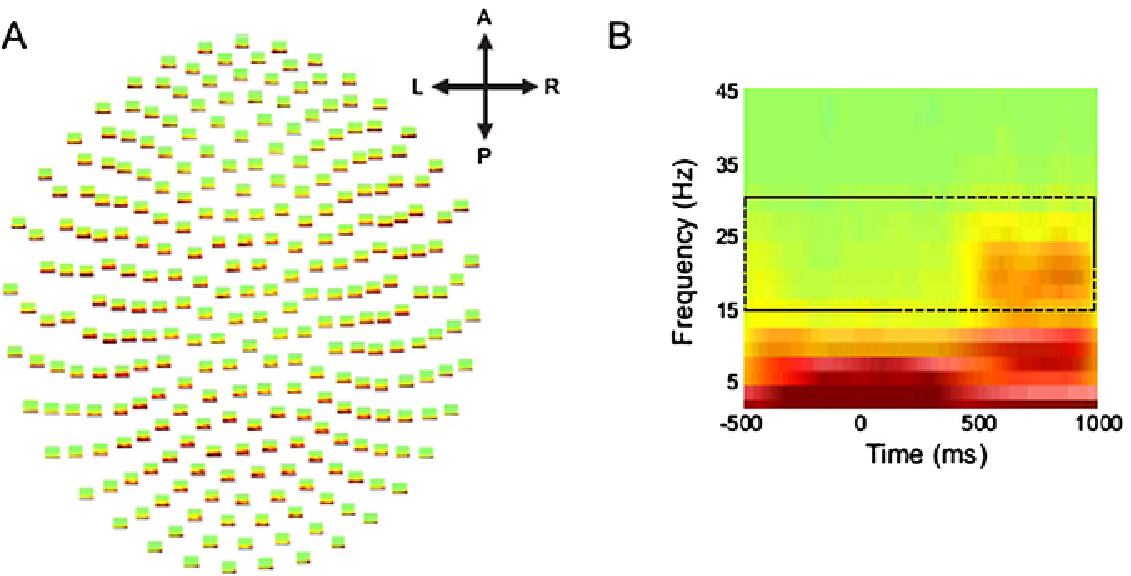}

\caption{Single-subject time--frequency data. (\textup{A}) shows the average TF data
across trials for the same subject shown in Figure~\protect\ref{fig1}. The~data are
plotted for 275 sensors across peristimulus time. (\textup{B})~shows the average TF
data across trials from a single sensor. For the subsequent analysis,
the TF maps were averaged across the 15--30 Hz frequency band for each
sensor. This band is shown in (\textup{B}) by the dotted lines.}\label{fig3}
\end{figure}

Time--frequency analyses of EEG/MEG recordings induce a 4D search-space,
at least two spatial dimensions, time and frequency [Figure~\ref{fig3}(A)].
Previously, we have shown that RFT can be applied to control for FWE
across 2D time--frequency SPMs, when the sensor-space of interest can be
defined a priori [Kilner,  Kiebel and Friston (\citeyear{KiKiFr2005})]. Here, we show that
when the frequency band of interest can be specified a priori
(often an easier specification), the resulting time-dependent modulation
of power in that frequency range can be treated in an identical fashion
to the ERF analysis described above. In other words, the 4D
data-features reduce to 3D, by averaging out frequency. In this example,
we averaged across frequency bins in the 15--30~Hz range [Figure~\ref{fig3}(B)].

\begin{figure}

\includegraphics{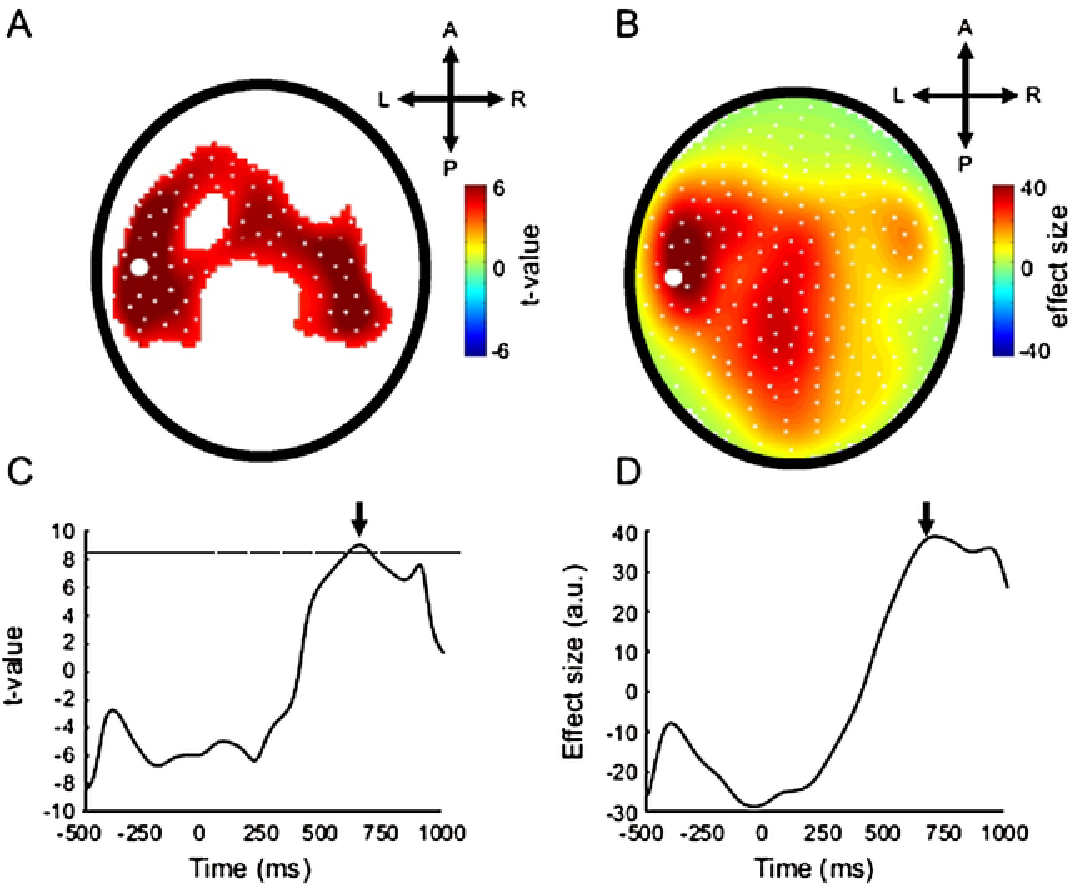}

\caption{SPM analysis of beta rebound. (\textup{A}) shows the
$\operatorname{SPM}(t)$, thresholded
at $p<0.001$ (uncorrected), showing where the effects were greater than
the mean. The~peak value within this cluster is significant at $p<0.05$
(corrected). The~most significant sensor is shown by a white circle.
(\textup{B})~shows the sensor-space map of effect size across all sensors at the time
where the SPM was maximal. (\textup{C})~shows the time course of the SPM from the
sensor shown in white in (\textup{A}). The~dashed lines show the FWE corrected
threshold for the Student's $t$-statistic at $p<0.05$. (\textup{D}) shows the
corresponding plot for the effect size. In both (\textup{C}) and (\textup{D}) the
arrow shows
the time where the SPM was maximal.}\label{fig4}
\end{figure}

The~space-time SPM for effects greater than the mean was calculated
using a one-sampled $t$-test as for the ERF analysis above. A large
spatial cluster contained a peak-value that was greater than the
threshold corrected for the entire search volume [$p<0.05$ corrected:
Figure~\ref{fig4}(A) and Table~\ref{tab3}]. The~peak occurred 560 ms after the button press
and was within central sensors over both the left and right hemispheres
[Figure~\ref{fig4}(A)---the sensor at the peak value is indicated by the white
circle---and Table \ref{tab1}]. Figure~\ref{fig4}(B) shows the average (nonstandardized)
effect-size across sensor-space at 560 ms. When comparing the
thresholded SPM [Figure~\ref{fig4}(A)] and the effect-size map [Figure~\ref{fig4}(B)], it is
clear that the sensor at the peak of the SPM is one of the sensors where
the effect is maximal [see also Figures~\ref{fig4}(C) and (D)]. Note that in the
effect-size map [Figure~\ref{fig4}(B)], the dipole field effects observed in Figure
\ref{fig2}(D) have the same sign, as the frequency decomposition renders the
data-features positive.

\section{Discussion} \label{sec:4}

We have illustrated how RFT can be employed to control FWE when making
statistical inference on continuous data, using movement-related MEG
responses that are continuous in space and time. We have not introduced
any novel methodology or statistical results. We have simply emphasized
the fact that established random field theory can be applied directly to
smooth, continuous data-features that conform to its assumptions. The~use of RFT may be particularly relevant for EEG and MEG data analysis,
which has to deal with data on manifolds that are not simple images and
may have a complicated geometry. In one sense, the contribution here is
to assert that one does not need novel methods for analyzing EEG and MEG
data-features, provided they exhibit continuity or smoothness properties
over connected vertices (or voxels). This is because inference is based
on topological quantities that do not depend on the coordinates or
geometry of those vertices (or voxels).

\begin{figure}[b!]
\begin{minipage}{\textwidth}
\begin{table}[H]
\caption{Statistical results of the time--frequency analysis in
\textup{SPM8}. See Table~\protect\ref{tab1} for details of the format}
\label{tab3}
\begin{tabular*}{\textwidth}{@{\extracolsep{\fill}}lccccc@{}}
\hline
\multicolumn{6}{@{}c@{}}{\textbf{Peak level}} \\
\hline
$\bolds{p}_\mathbf{FWE\mbox{\textbf{-}}corr}$ & $\bolds{p}_\mathbf{FDR\mbox{\textbf{-}}corr}$ & $\bolds{t}$ & $\bolds{Z}$ & $\bolds{p}_\mathbf{uncorrected}$ & \textbf{Time (ms)}\\
\hline
0.033 & 0.001 & 9.05 & 4.75 & 0.000 & 560\\
\hline
\end{tabular*}
\tabnotetext[]{tm}{Statistics: $p$-values restricted to the entire search volume.
Height threshold: $T = 4.02$, $p =0.001$ (0.966); Degrees of freedom $= [1.0, 11.0]$;
Extent threshold: $k = 0$ bins, $p = 1.000$ (0.966); Smoothness $\mathrm{FWHM}= 13.3$ 13.4 17.1 \{bins\};
Expected bins peir cluster, $\langle k\rangle  = 175.646$; Search vol.: 1,808,083 bins; 149.4 resels;
Expected number of clusters, $\langle c\rangle  = 3.40$;
Expected false discovery rate, ${\leq}0.01$.}
\end{table}
\end{minipage}
\end{figure}

\subsection{Random field theory assumptions} \label{sec:4.1}

One of the assumptions of RFT is that the error fields conform to a good
lattice approximation of an underlying random field [Worsley (\citeyear{Wo2007})]. In
other words, the underlying random field must be sampled sufficiently
densely so as to be able to estimate the smoothness of the underlying
random field. In practice, this means that one must ensure that there is
a sufficiently high sampling of EEG/MEG signals across the dimensions of
interest. In the examples presented above these would be sensor-space
and time. For high-density EEG and MEG, with a large number of sensors
covering the scalp surface, this requirement is clearly met. However,
care must be taken if one wanted to adopt this approach for sparsely
sampled EEG/MEG data, as this assumption may be violated. In such cases,
it is noteworthy that RFT can accommodate any $D$-dimensional
search-spaces and can therefore be applied to time-courses from a single
sensor or some summary over sensors [cf. Carbonell et al. (\citeyear{CaGaVaWoBiDaBoPa2004})].

RFT requires that the random fields are multivariate Gaussian with a
differentiable correlation function [Worsley (\citeyear{Wo2007})]. The~correlations
do not have to be stationary when controlling the FWE. This means that
any nonstationarity that is induced by flattening a manifold in 3D-space
to a 2D sensor-space does not violate the assumptions of RFT. RFT
procedures can be used to characterize other topological features of the
SPM, namely, the extent and number of clusters. When using RFT to
control the FWE rate for cluster-size inference, one effectively
measures the size of each cluster in resels, which accommodates local
smoothness. It should be noted, however, that the current \textit{SPM8}
implementation does not do this properly (for reasons of computational
expediency) and that inference on cluster-size assumes isotropic
smoothness, which is usually induced by smoothing the data [see also
Salmond et al. (\citeyear{SaAsVaCoGaFr2002})]. In the absence of this smoothing, better
approximations are available (e.g., \textit{SurfStat}).

Topological inference enables the control of FWE rate across a search
volume when making statistical inferences. Therefore, the approach can
be adopted in any situation in which one would normally perform
parametric statistical tests, such as a $t$- or $F$-test. When parametric
statistical tests cannot be used, for example, when the errors are not
normally distributed, the requisite null distribution of the maximal
statistic can be estimated using nonparametric procedures. Nonparametric
methods have been used to make statistical inference on both MEG
source-space data [Singh, Barnes and Hillebrand (\citeyear{SiBaHi2003}); Pantazis et al. (\citeyear{PaNiBaLe2005})] and on
clusters in space, time and frequency [Maris and Oostenveld (\citeyear{MaOo2007}), see
also \url{http://www.ru.nl/neuroimaging/fieldtrip/}]. However, the
analytic and closed form expressions provided by RFT are based on
assumptions that, if met, render it more powerful or sensitive than
equivalent nonparametric approaches [Howell (\citeyear{Ho1997})]. Furthermore, with
appropriate transformations [e.g., Kiebel,  Tallon-Baudry and Friston (\citeyear{KiTaFr2005})] and post hoc smoothing, it is actually quite difficult to contrive situations
where the errors are not multivariate Gaussian (by the central limit
theorem) and violate the assumptions of RFT.

\subsection{Topological inference in space and time}

In this note we have shown how RFT can be used to solve the multiple
comparisons problem that besets statistical inference using EEG/MEG
data. This approach has several advantages. First, it avoids ad hoc or selective characterization of data inherent in conventional
approaches that use averages over pre-specified regions of search-space.
Second, inference is based on $p$-values that are adjusted for multiple
nonindependent comparisons, even when dependencies have a complicated
form. Third, this adjustment is based explicitly on the search-space,
giving the researcher the latitude to restrict the search-space to the
extent that prior information dictates. In short, topological inference
enables one to test for effects without knowing where they are in space
or time. This may be useful, as it could disclose effects that hitherto
may have gone untested, for example, small effects that are highly
reproducible. However, as we intimated above, the neurophysiological
interpretation of significant effects must be considered in light of the
quantitative estimates one is making an inference about. In conventional
analyses, prior knowledge about the effects of interest is used to
average the data to finesse the multiple comparisons problem. With
topological inference, these a priori constraints are used to
reduce the search-space and adjust the $p$-values of the SPM within this
reduced volume [Figures~\ref{fig2}(C)--(E)]. For example, if one is interested in
modulations of the N170, one could reduce the search-space to a time
window spanning 140--200 ms post-stimulus. If, in addition, one had
predictions about where this effect should be observed in sensor or
source space, then one could reduce the search volume even further. The~advantage of this, over conventional averaging, is that inference may be
more sensitive, as it pertains to peak responses that are necessarily
suppressed by averaging.

\subsection{Topological FDR} \label{sec:4.4}

We have focused on the use of random field theory for controlling the
false positive rate of topological features in statistical maps.
However, there is a growing interest in applying the same ideas to
control false discovery rate [Benjamini and Hochberg (\citeyear{BeHo1995})]. Crucially,
topological FDR controls the expected false discovery rate (FDR) of
features (such as peaks or excursion sets), as opposed to simply
controlling the FDR of point tests (e.g., Student's \mbox{$t$-tests} at each
voxel or vertex). This is because FDR procedures in imaging can be
problematic and lead to capricious inference [Chumbley and Friston
(\citeyear{ChFr2009})]. The~reason is that most image analysis deals with signals that
are continuous (analytic) functions of some support; for example, space
or time. In the absence of bounded support, the false discovery rate
must be zero. This is because every discovery is a true discovery, given
that the signal is (strictly speaking) everywhere. Crucially, one can
finesse this problem by inferring on the topological features of the
signal. For example, one can assign a $p$-value to each local maximum in
an SPM using random field theory and identify an adaptive threshold that
controls false discovery rate, using the Benjamini and Hochberg procedure
[Chumbley et al. (\citeyear{ChWoFlFr2010})]. This is called topological
FDR and provides a natural complement to conventional FWE control. The~notion of topological FDR was introduced in a paper that was the last to
be co-authored by Keith Worsley, shortly before his death.

\subsection{Conclusion}

We have illustrated how topological inference can be applied to EEG/MEG
data-features that vary as a smooth function of frequency, time or space
and have stressed the generality of this application. These procedures
have a number of advantages: (i) They require no a priori
specification of where effects are expressed, (ii) inferences are based
on $p$-values that are adjusted for multiple comparisons of continuous
and highly correlated data-features and (iii) these inferences are
potentially more sensitive than tests on regional averages. One might
anticipate that the advances made by Keith Worsley will find new and
important domains of application as people start to appreciate the
generality and simplicity of his legacy.

\section*{Acknowledgments}
We would
like to thank Stefan Kiebel for helpful comments on an earlier version
of this manuscript.

\printaddresses

\end{document}